# Thermal conductivity of intrinsic semiconductor at elevated temperature: role of four-phonon scattering and electronic heat conduction

Xiaokun Gu,[1#*] Shouhang Li,[2#] and Hua Bao[2*]

[1]Institute of Engineering Thermophysics, School of Mechanical Engineering, Shanghai Jiao Tong University, Shanghai 200240, China
[2]University of Michigan-Shanghai Jiao Tong University Joint Institute, Shanghai Jiao Tong University, Shanghai 200240, China
[#]Contributed equally to this work
[*]Email: Xiaokun.Gu@sjtu.edu.cn, Hua.Bao@sjtu.edu.cn

**Abstract**

While using first-principles-based Boltzmann transport equation approach to predict the thermal conductivity of crystalline semiconductor materials has been a routine, the validity of the approach is seldom tested for high-temperature conditions. Most previous studies only focused on the phononic contribution, and neglected the electronic part. Meanwhile, the treatment on phonon transport is not rigorous as a few ingredients, such as four-phonon scatterings, phonon renormalization and thermal expansion, are ignored. In this paper, we present a Boltzmann transport equation study on high-temperature thermal conduction in bulk silicon by considering the effects of both phonons and electrons, and explore the role of the missing parts in the previous studies on the thermal conductivity at elevated temperature. For the phonon transport, four-phonon scattering is found to considerably reduce the thermal conductivity when the temperature is larger than 700 K, while the effects of phonon renormalization and thermal expansion on phononic thermal conductivity are negligible. Bipolar contribution to the electronic thermal conductivity calculated from first-principles is implemented for the first time. More than 25% of heat is shown to be conducted by electrons at 1500 K. The computed total thermal conductivity of silicon faithfully reproduces the measured data. The approach presented in this paper is expected to be applied to other high-temperature functional materials, and the results could serve as benchmarks and help to explain the high-temperature phonon and electron transport phenomena.

**I. Introduction**

Thermal conductivity, $\kappa$, is a key material property that describes the relation between heat flux, $\mathbf{J}$, and temperature gradient, $\nabla T$, within the material through Fourier's law of heat conduction, $\mathbf{J} = -\kappa \cdot \nabla T$. Intensive efforts had been devoted to understand the microscopic mechanisms of heat conduction and to predict the thermal conductivity of materials in the past few decades. While recent advances have made it possible to predict both lattice and electron contributions to the thermal conductivity through first-principles-based calculations if the temperature is not high,[1-2] a theoretical approach that captures the all essential transport characteristics of various energy carriers at high temperatures is still lacking. The knowledge of thermal transport in crystalline solids under high temperatures is of great importance in designing thermal barrier coatings,[3] developing thermoelectric materials,[4] finding solutions for heat dissipation in high power electronics[5-7] and understanding the planetary evolution.[8]

Remarkable progress in predicting phononic thermal conductivity from first-principles,[9] including molecular dynamics (MD) simulations and Peierls-Boltzmann transport equation (PBTE) calculations, has been achieved in the past ten years. There have been a few different approaches proposed to integrate first-principles with MD simulations so as to satisfactorily describe the phonon dispersion and anharmonic effects.[10-11] However, due to the severe computational costs for first-principles-based MD simulations, the simulation domains are usually limited to hundreds of atoms, which might lead to strong size effects.[12-13] Alternatively, in PBTE-based methods, phonon transport is explored in reciprocal space and the artificial size effects could be avoided. Using interatomic force constants (IFCs) extracted from first-principles calculations as inputs and solving the PBTE, the obtained thermal conductivity values usually agree reasonably well with experimental results for a wide range of materials including common semiconductors [14-15] and thermoelectric materials.[16] Despite the success of the first-principles-based PBTE method, the standard implementation of this approach might lead to considerable uncertainties when applied to high temperature materials, as some important ingredients that affect both phonon and electron transport at high temperatures are ignored.

In most PBTE studies, only the lowest order contributions to the anharmonicity (three-phonon scatterings) are accounted for.[17-19] While the three-phonon scatterings are believed to be dominant at low temperatures, higher-order phonon-phonon scatterings are expected to become non-negligible with the increase of the temperature.[20] Another outcome due to high temperature is the change of IFCs, but most previous PBTE calculations relied on the IFCs extracted at 0 K, except a few very recent works.[21] Since the IFCs are directly linked with phonon dispersion and phonon scattering rates, it is necessary to include the temperature dependent IFCs in PBTE calculations. The other factor might affect the thermal conduction due to phonons is the thermal expansion, as the IFCs and correspondingly the phonon properties should be lattice-constant-dependent.[22] Despite the fact that these high-temperature effects have been explored in some studies, separately, the studies including all these ingredients are still scare, except very recent ones.[23-24]

In addition to the contributions of phonons to the total thermal conductivity, electrons are the other main heat carriers. While there is no need to mention the importance of electrons to the heat conduction in metals and doped semiconductors, the roles of electrons could be significant for intrinsic semiconductors under elevated temperatures due to the increased intrinsic carrier concentration. It was noticed that the theoretical evaluation of thermal conductivity only considering phonons does not match experimental results at high temperature. For example, several decades ago, Glassbrenner and Slack[25] found that based on an empirical model for phononic thermal conductivity the predicted thermal resistance of silicon would be overestimated in high temperature range (>1000 K), though many rough approximations makes the accuracy of the model they employed far from satisfactory. Li *et al.*[26] also observed the difference between the phononic thermal conductivity of $Mg_2Sn$ from their first-principles BTE calculations and the total thermal conductivity from measurements when the temperature is higher than 400 K. To account for the electronic heat conduction in intrinsic semiconductors at high temperatures, a few models were proposed. The most widely used method is Wiedemann-Franz law,[27-28] i.e. the electronic thermal conductivity $\kappa_e$ is established from electrical conductivity $\sigma$ with the equation $\kappa_e = L\sigma T$, with $L$ and $T$ are Lorenz number and temperature, respectively. The issue of this method is the uncertainty in the estimation of Lorenz number $L$. Although the widely used Sommerfeld value $L_0 = \frac{\pi^2}{3}\left(\frac{k_B}{e}\right)^2 = 2.44 \times 10^{-8} \, W\Omega/K^2$ is quite accurate for metals at high temperatures,[29] recent studies show it has strong variation in semiconductors.[16, 30-31] The semi-empirical deformation potential model [32-33] is also widely used to evaluate the charge carrier energy dependent electron-phonon scattering rate and then obtain the electrical transport properties. Nevertheless, the assumptions, like the long-wavelength limit assumption for acoustic phonon and parabolic band for electron, limit its accuracy in the estimation of electrical transport properties of semiconductors[34]. Recently, a rigorous first-principles method was promoted and has been widely used to predict electrical transport properties. Such method is based on Fermi's golden rule employing Maximally Localized Wannier Function

(MLWF) [35] interpolation technique which makes the high accuracy mode-by-mode calculation available.[36] Despite the success of this method in studying the electrical and thermal transport properties of semiconductors[2, 16, 34] and metals,[37-39] it has not been applied to study the electronic heat conduction in intrinsic semiconductors at high temperature.

In this paper, we take bulk intrinsic silicon as a typical example of crystalline materials and conduct a thorough investigation on its thermal conductivity of bulk intrinsic silicon under a wide temperature range (from 100 K to 1500 K) based on the first-principles-based Boltzmann transport equation approach. Compared with previous studies, which mainly focused on phononic thermal conductivity, both the contributions from phonons and electrons to the total thermal conductivity are taken into account in this study. The effects of four-phonon scatterings, the temperature dependent IFCs and thermal expansion on phononic thermal conductivity are examined. We also explore the roles of electronic thermal conductivity, including the polar and bipolar contributions. With our approach, the obtained total thermal conductivity of silicon can agree with the measured data well within the whole temperature range explored here. The framework presented could be applied to other crystalline materials, and paves the path to fully understand the thermal conductivity of high-temperature functional materials.

## II. Numerical methods

Since silicon is a semiconductor, the heat is carried by phonons if the temperature is not high. However, the charge carrier concentration exponentially increases with the temperature, which could result in none negligible contribution to heat conduction at high temperature. Including the contributions from both phonons and electrons, the thermal conductivity of silicon could be written as

$$\kappa = \kappa_{\text{ph}} + \kappa_{\text{e}}, \qquad (1)$$

where $\kappa_{\text{ph}}$ and $\kappa_{\text{e}}$ are the phononic and electronic thermal conductivity, respectively. Here, we apply the Boltzmann transport equation to determine the phononic and electronic thermal conductivity based on the inputs from first-principles calculations. In Sec. A, the formalism of PBTE to determine the phononic thermal conductivity is briefly reviewed, followed by the discussion on the method to extract the IFCs under finite temperatures. In Sec. C, the theory to compute the electronic thermal conductivity is introduced.

### A. Phononic thermal conductivity from PBTE

When there is no temperature gradient in the crystal, the phonon population function obeys the Bose-Einstein distribution, $n_\lambda^0 = 1/\left[\exp(\hbar\omega_\lambda / k_B T) - 1\right]$, where $\lambda$ represents the *s*-th phonon mode at **q** in the first Brillouin zone, $\omega$ is the phonon frequency, $\hbar$, $k_B$ and $T$ are the reduced Planck's constant, Boltzmann constant and temperature, respectively. Once a temperature gradient is applied, e.g., the phonon population function becomes deviating from the equilibrium one and could be expressed as $n_\lambda = n_\lambda^0 + n_\lambda^0(n_\lambda^0 + 1)\mathbf{F}_\lambda \cdot \nabla T$, with the unknown deviation function $\mathbf{F}_\lambda$. Supposing the deviation function for each phonon mode is known, the heat flux within the material can be computed by summing up the contribution from all modes through[40-41]

$$\mathbf{J} = \sum_\lambda \hbar\omega_\lambda \mathbf{v}_\lambda n_\lambda = \left(\sum_\lambda \hbar\omega_\lambda n_\lambda^0(n_\lambda^0 + 1)\mathbf{v}_\lambda \mathbf{F}_\lambda\right) \cdot \nabla T. \tag{2}$$

Thus, according to the Fourier's law, the thermal conductivity tensor is

$$\kappa^{\alpha\beta} = \sum_\lambda \hbar\omega_\lambda n_\lambda^0(n_\lambda^0 + 1) v_\lambda^\alpha F_\lambda^\beta. \tag{3}$$

In order to determine the non-equilibrium phonon population function, or equivalently the deviation function, the PBTE is solved. Since the heat conduction in silicon is isotropic, we only consider the *x* direction. Under the relaxation time approximation, the PBTE is written as[33]

$$v_\lambda^x \frac{\partial T}{\partial x} \frac{\partial n_\lambda^0}{\partial T} = \frac{n_\lambda^0 - n_\lambda}{\tau_\lambda}, \tag{4}$$

where $\tau_\lambda$ is the phonon lifetime. Solving the PBTE and substitute the obtained $F_\lambda^x$ to Eq. (3), the expression of thermal conductivity is further simplified to

$$\kappa^{xx} = \sum_\lambda \frac{\hbar}{k_B T^2} n_\lambda^0(n_\lambda^0 + 1) \omega_\lambda^2 (v_\lambda^x)^2 \tau_\lambda. \tag{5}$$

Assuming the validity of Matthiessen's rule, the phonon lifetime can be written as

$$\frac{1}{\tau_{\mathbf{q}s}} = \frac{1}{\tau_{\mathbf{q}s}^{3\text{ph}}} + \frac{1}{\tau_{\mathbf{q}s}^{4\text{ph}}} + \frac{1}{\tau_{\mathbf{q}s}^{\text{iso}}}, \tag{6}$$

where $\tau_{\mathbf{q}s}^{3\text{ph}}$, $\tau_{\mathbf{q}s}^{4\text{ph}}$ and $\tau_{\mathbf{q}s}^{\text{iso}}$ are the phonon lifetimes due to three-phonon, four-phonon and phonon-isotope scatterings. From the Fermi's golden rule, the phonon lifetime of three-phonon and four-phonon scatterings of mode **q***s* are computed through

$$\frac{1}{\tau_{\mathbf{q}s}^{3ph}} = 2\pi \sum_{\mathbf{q}'s'} \sum_{s''} |V_3(\mathbf{q}s, \mathbf{q}'s', \mathbf{q}''s'')|^2 \cdot \tag{7}$$

$$\left[\left(n_{\mathbf{q}'s'}^0 - n_{\mathbf{q}''s''}^0\right)\delta(\omega_{\mathbf{q}s} + \omega_{\mathbf{q}'s'} - \omega_{\mathbf{q}''s''}) + \frac{1}{2}\left(1 + n_{\mathbf{q}'s'}^0 + n_{\mathbf{q}''s''}^0\right)\delta(\omega_{\mathbf{q}s} - \omega_{\mathbf{q}'s'} - \omega_{\mathbf{q}''s''})\right]$$

and

$$\frac{1}{\tau_{qs}^{4ph}} = 2\pi \sum_{q's'} \sum_{q''s''} \sum_{s'''} |V_4(qs, q's', q''s'', q'''s''')|^2 \cdot \qquad (8)$$

$$\left[ \frac{1}{2} \frac{n^0_{q's'} n^0_{q''s''} (n^0_{q'''s'''} + 1)}{(n^0_{qs} + 1)} \delta(\omega_{qs} + \omega_{q's'} + \omega_{q''s''} - \omega_{q'''s'''}) \right.$$

$$+ \frac{1}{2} \frac{(n^0_{q's'} + 1) n^0_{q''s''} n^0_{q'''s'''}}{n^0_{qs}} \delta(\omega_{qs} + \omega_{q's'} - \omega_{q''s''} - \omega_{q'''s'''})$$

$$+ \left. \frac{1}{6} \frac{n^0_{q's'} n^0_{q''s''} n^0_{q'''s'''}}{n^0_{qs}} \delta(\omega_{qs} - \omega_{q's'} - \omega_{q''s''} - \omega_{q'''s'''}) \right]$$

where $V_3$ and $V_4$ are the three-phonon and four-phonon scattering matrix elements. The phonon scattering matrix elements are determined by the third-order and fourth-order anharmonic force constants. The expressions for the scattering matrix elements can be documented in Ref. [42]. The phonon-isotope phonon lifetimes are evaluated by the Tamura theory,[43]

$$\frac{1}{\tau_{qs}^{4ph}} = \frac{\pi}{2} \omega_{qs} \omega_{q's'} \sum_{\tau} g_{\tau} \left| \sum_{\alpha} (\varepsilon^{\tau\alpha}_{qs})^* \varepsilon^{\tau\alpha}_{q's'} \right|^2 \delta(\omega_{qs} - \omega_{q's'}), \qquad (9)$$

where $g_\tau$ is the mass variance parameter for the basis atom $\tau$, and $\varepsilon$ is the eigenvector of the phonon mode. In the naturally occurring Si, $g_{Si} = 2.01 \times 10^{-4}$.

### B. Temperature dependent interatomic force constants

In order to evaluate the lattice thermal conductivity, phonon dispersion and phonon lifetime have to be computed. The phonon dispersion is determined by the harmonic force, while the anharmonic force constants are required to calculate the phonon lifetimes. Typically, these IFCs are defined through[44]

$$H = K + V = K + E_0 + \frac{1}{2!} \sum_{ij} \sum_{\alpha\beta} \phi^{\alpha\beta}_{ij} u^\alpha_i u^\beta_j + \frac{1}{3!} \sum_{ijk} \sum_{\alpha\beta\gamma} \psi^{\alpha\beta\gamma}_{ijk} u^\alpha_i u^\beta_j u^\gamma_k \qquad (10)$$

$$+ \frac{1}{4!} \sum_{ijkl} \sum_{\alpha\beta\gamma\theta} \chi^{\alpha\beta\gamma\theta}_{ijkl} u^\alpha_i u^\beta_j u^\gamma_k u^\theta_l + \cdots$$

where $H$, $K$ and $V$ are the Hamiltonian, kinetic energy and potential energy of the crystal; $I$ and $\alpha$ mean the $\alpha$ direction of atom $I$; $u$ is the atomic displacement away from the equilibrium position; $\phi$, $\psi$ and $\chi$ are the second-order harmonic, third-order and fourth-order anharmonic force constants, which are regarded as the derivatives of the potential with respect to the atomic displacement; $E_0$ is the potential energy when $u = 0$. Ideally, when the temperature is low (the amplitudes of atomic displacements are small), the harmonic terms are dominating compared with the anharmonic terms. The vibrational pattern of the atoms in the crystal is governed by the dynamical equation, where the harmonic force constants appear. By diagonalizing the corresponding dynamical matrix, the phonon dispersion and eigenvectors are solved. At higher-than-zero temperatures, due to the anharmonicity of the lattice, the phonon dispersion relation should be deviating from the static limit, which is usually called phonon renormalization. Since the strengths of phonon scatterings are partially determined by the phonon scattering phase space, the effects of phonon renormalization on the thermal conductivity cannot be ignored in highly anharmonic crystals and/or under high temperatures. There have been a few approaches proposed to treat phonon renormalization,[45]. Among them, TDEP was found to work well for predicting the thermal conductivity of a variety of crystalline materials, including $Bi_2Te_3$,[21] $PbTe$[46] and graphene.[23] Concerning its efficiency and its compatibility with convectional direct method to extract force constants, in this work we adopt TDEP to extract temperature dependent force constants.

At high temperatures, an effective harmonic potential is used to evaluate the potential surface, through which the temperature dependent phonon dispersion could be obtained. The effective potential is expressed as

$$V \approx \tilde{E}_0 + \frac{1}{2!}\sum_{ij}\sum_{\alpha\beta} \tilde{\phi}_{ij}^{\alpha\beta} u_i^\alpha u_j^\beta, \tag{11}$$

where $\tilde{E}_0$ and $\tilde{\phi}$ are the effective equilibrium potential energy and harmonic force constants. According to Eq. (10), the second-order effective harmonic force constants can be extracted by fitting the displacement-atomic force data through

$$F_i^\alpha = -\frac{\partial V}{\partial u_i^\alpha} \approx -\sum_j \sum_\beta \hat{\phi}_{ij}^{\alpha\beta} u_j^\beta. \tag{12}$$

Comparing the real potential surface and effective one, the residue leads to perturbations to the effective harmonic potential. Here, this part is analytically written as the polynomials of atom displacements with the coefficients ($\hat{\psi}$, $\hat{\chi}$) as the temperature dependent anharmonic force constants, i.e.,

$$V - \tilde{E}_0 - \frac{1}{2!}\sum_{ij}\sum_{\alpha\beta} \tilde{\phi}_{ij}^{\alpha\beta} u_i^\alpha u_j^\beta = \frac{1}{3!}\sum_{jk}\sum_{\beta\gamma} \hat{\psi}_{ijk}^{\alpha\beta\gamma} u_i^\alpha u_j^\beta u_k^\gamma + \frac{1}{4!}\sum_{jkl}\sum_{\beta\gamma\theta} \hat{\chi}_{ijkl}^{\alpha\beta\gamma\theta} u_i^\alpha u_j^\beta u_k^\gamma u_l^\theta. \tag{13}$$

The coefficients are extracted by fitting the following relation

$$F_i^\alpha - \left(-\sum_j\sum_\beta \hat{\phi}_{ij}^{\alpha\beta} u_j^\beta\right) = -\frac{1}{2!}\sum_{jk}\sum_{\beta\gamma} \hat{\psi}_{ijk}^{\alpha\beta\gamma} u_j^\beta u_k^\gamma - \frac{1}{3!}\sum_{jkl}\sum_{\beta\gamma\theta} \hat{\chi}_{ijkl}^{\alpha\beta\gamma\theta} u_j^\beta u_k^\gamma u_l^\theta. \tag{14}$$

To explore the roles of temperature dependent force constants, we also extract the force constants at the static limit by the finite difference scheme, in which one, two or three atoms are displaced along the

Cartesian coordinates by a small distance.

### C. Polar and bipolar effect of intrinsic semiconductor

In intrinsic semiconductor, there are two categories of charge carriers, namely electrons and holes that can conduct heat and electricity. The electronic thermal conductivity is made up of two components, the polar part ($\kappa_p$) and bipolar part ($\kappa_b$). The polar heat flux is generated by the diffusion process of the same charge carrier category (electron or hole) between the hot side and cold side. Meanwhile, the hot electron and hole can have recombination at cold side. Then large amount of heat can be released due to the large energy difference (~ band gap energy $\epsilon_g$) between electron and hole which is denoted as bipolar heat flux. The total electronic thermal conductivity can be expressed as [47]

$$\kappa_e = \kappa_{ep} + \kappa_{eb} = \kappa_{\text{ep},n} + \kappa_{\text{ep},p} + \frac{\sigma_n \sigma_p}{\sigma_n + \sigma_p}(\alpha_n - \alpha_p)^2 T, \tag{15}$$

where $\kappa_{\text{el},n} + \kappa_{\text{el},p}$ is the polar component and the third term is the bipolar component. The subscript $n$ and $p$ represent electron and hole carrier, respectively. $\kappa_e$ is electronic thermal conductivity, $\sigma$ is electrical conductivity, $\alpha$ is Seebeck coefficient. The Seebeck coefficient is positive for holes and negative for electrons. For intrinsic semiconductor, the magnitudes of $\alpha_n$ and $\alpha_p$ are comparable with different signs which will make the $(\alpha_n - \alpha_p)$ term large. Therefore, at high temperature, the bipolar term contribution can be significantly larger than that of polar term.

Based on the Onsager reciprocal relations,[48] electrical conductivity $\sigma_n(\mu, T)$, Seebeck coefficient $S_n(\mu, T)$ and electronic thermal conductivity $\kappa_{\text{ep},n}(\mu, T)$ for electron can be expressed as following

$$\begin{cases} \sigma_n(\mu,T) = \frac{1}{N_k \Omega} \sum_{nk} -e^2 v_{nk}^2 \tau_{nk}(\mu,T) \frac{\partial f(\varepsilon_{nk},\mu,T)}{\partial \varepsilon_{nk}} \\ S_n(\mu,T) = -\frac{1}{eT} \frac{\sum_{nk} (\varepsilon_{nk}-\mu) v_{nk}^2 \tau_{nk}(\mu,T) \frac{\partial f(\varepsilon_{nk},\mu,T)}{\partial \varepsilon}}{\sum_{nk} v_{nk}^2 \tau_{nk}(\mu,T) \frac{\partial f(\varepsilon_{nk},\mu,T)}{\partial \varepsilon_{nk}}} \\ \kappa_{\text{ep},n}(\mu,T) = \frac{1}{N_k \Omega} \sum_{nk} -\frac{(\varepsilon_{nk}-\mu)}{T} v_{nk}^2 \tau_{nk}(\mu,T) \frac{\partial f(\varepsilon_{nk},\mu,T)}{\partial \varepsilon} - TS^2(\mu,T)\sigma(\mu,T) \end{cases}, \tag{16}$$

where $N_k$ is the total number of **k**-points, $\Omega$ is the volume of the unit cell, $e$ is the elementary charge and $v_{nk}$ is the group velocity of electron denoted with electronic wave vector **k** and conduction band index $n$. $f(\varepsilon_{m\mathbf{k+q}}, \mu, T)$ is equilibrium electron distribution function, namely Fermi-Dirac distribution, with $\mu$ Fermi level and $\varepsilon_{nk}$ electron energy. $\tau_{nk}(\mu, T)$ is the charge carrier relaxation time. It is mode dependent resulting from the electron-phonon scattering within Fermi's golden rule, which is the kernel in the calculation of electrical transport properties.

The electron scattering rate $1/\tau_{n\mathbf{k}}$ is related to the imaginary part of the lowest-order of electron self-energy,[36] i.e. $1/\tau_{nk} = 2/\hbar \mathrm{Im}\Sigma_{n\mathbf{k}}$. The electron self-energy $\Sigma_{n\mathbf{k}}$ can be expressed as [36]

$$\Sigma_{m\mathbf{k}}(\omega, T) = \sum_{ns} \frac{d\mathbf{q}}{\Omega_{BZ}} |g_{mn,s}(\mathbf{k}, \mathbf{q})|^2 \times \left[ \frac{n_{\mathbf{q}s}(T)+f_{m\mathbf{k}+\mathbf{q}}(T)}{\omega-(\varepsilon_{m\mathbf{k}+\mathbf{q}}-\mu)+\omega_{\mathbf{q}s}+i\delta} + \frac{n_{\mathbf{q}s}(T)+1-f_{m\mathbf{k}+\mathbf{q}}(T)}{\omega-(\varepsilon_{m\mathbf{k}+\mathbf{q}}-\mu)-\omega_{\mathbf{q}s}+i\delta} \right] \quad (17)$$

$\Omega_{BZ}$ is the volume of the first Brillouin zone. $\delta$ is a small positive parameter to ensure the numerical stability. The electron-phonon coupling matrix element[49] $g_{mn}^s(\mathbf{k},\mathbf{q})$ quantifies the strength of electron-phonon scattering process, which describes the electron with initial state of $n\mathbf{k}$ approaches to final state of $m\mathbf{k}+\mathbf{q}$ with the perturbation of phone mode of $\mathbf{q}s$ and it is denoted as

$$g_{mn}^s(\mathbf{k}, \mathbf{q}) = \langle \psi_{m\mathbf{k}+\mathbf{q}} | \nabla_{\mathbf{q}s} V | \psi_{n\mathbf{k}} \rangle \quad (18)$$

where $\psi_{n\mathbf{k}}$ and $\psi_{m\mathbf{k}+\mathbf{q}}$ are two different electron states. $\nabla_{\mathbf{q}s} V$ denotes the first-order derivative of the Kohn-Sham potential with respect to the phonon displacement.

## III. Simulation details

We performed calculations within density-functional theory (DFT) on silicon with a relaxed lattice constant of 5.43 Å using the generalized gradient approximation of Perdew, Burke, and Ernzerhof for solids (PBEsol) [50] and a planewave basis with the QUANTUM ESPRESSO package.[51] An energy cutoff of 45 Ry was employed to obtain the ground-state charge density and band structure. For phononic thermal conductivity calculation, the force constants are computed by fitting the force-displacement data in supercells. The supercells consist of 3×3×3 cubic conventional cells with 216 atoms.

To perform zone integration for computing phonon scattering rats and phononic thermal conductivity, the first Brillouin zone is decomposed to $N \times N \times N$ q-point meshes. It has been recognized the obtained thermal conductivity depends on the value of $N$. To eliminate the dependence of the q-point mesh on the results, we linearly fit the relation between the thermal conductivity data computed with $N$ = 16, 20, 24 and 28 and $1/N$, which provides the mesh-independent thermal conductivity by letting $1/N$ = 0. In addition, the computational cost for computing the four-phonon scattering rates increases rapidly with $N$. In order to make the four-phonon calculation affordable, we adopt the numerical technique, through which the four-phonon scattering rates of the phonon modes in the mesh with large $N$ are extrapolated with the scattering rates from a mesh with $N$ = 12.[52] By using the extrapolation method, the computational cost is within serval hundred CPU hours.

For the electrical transport properties calculation, a 12×12×12 Monkhorst-Pack **k**-point mesh was used for both the self-consistent and non-self-consistent field calculations. The lattice dynamical properties were predicted based on Density-Functional Perturbation Theory[53-54] (DFPT) method, in which a 6×6×6 **q**-point mesh was implemented. The threshold for phonon calculation was set as $10^{-14}$ Ry.

The prediction of electron band gap from DFT is 0.75 eV, which is smaller than experimental value and due to the inherent defect of DFT.[55-56] However, the electron transport properties are strongly related to the band gap which is significantly affected by the temperature[57]. Due to the difficulty of predicting band gap at high temperature from DFT, Varshni's equation[58] and Thurmond's data[59] for silicon $\epsilon_g(T) = 1.17 - 4.73 \times 10^{-4} T^2/(T + 636)$ eV was employed to obtain the temperature dependent band gap. Then, we shift the conduction band with respect to the electron energy band gap based on the rigid band approximation.[60]

The Electron-Phonon Wannier (EPW) package[36] was employed to calculate the electron-phonon scattering rate. We use 40×40×40 dense **q**-point mesh and 100×100×100 **k**-point mesh by weighting the convergence and computation cost.

## IV. Results and discussion

### A. Phononic thermal conductivity

We first calculate the phononic thermal conductivity of the naturally occurring silicon (with phonon-isotope scatterings) between 100 K and 1500 K by accounting for three-phonon scatterings as the only phonon-phonon scattering mechanism and using the IFCs at 0 K. Such calculations have been conducted by many groups. These data are serving as benchmarks for other calculations where more ingredients relevant to high-temperature transport are included. The obtained results computed are presented in Fig. 1(a) as solid squares, along with the measured data taken from Ref. [61], [25] and [62] (lines). The temperature dependence of the calculated thermal conductivity agrees reasonably well with the experimental data for the temperatures under 500 K. However, when the temperature is elevated, noticeable difference can be clearly identified. For example, at 900 K, the calculated phononic thermal conductivity is 43 W/mK, 20% larger than the measured data. With the temperature further enhanced, the difference between the calculation and measurement becomes smaller. The origins of the discrepancy will be discussed below.

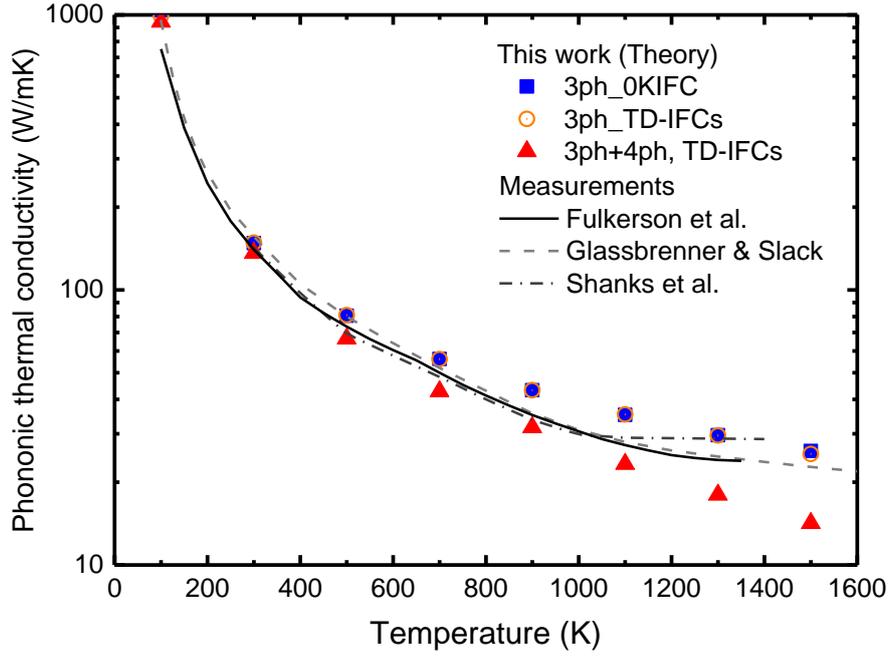

Figure 1. Phononic thermal conductivity of bulk silicon as a function of temperature. The experimental data is extracted from the work of Fulkerson *et al.* [61] (solid line), Glassbrenner & Slack[25] (dash line) and Shanks *et al.*[62] (dash dot line), respectively.

To explore the role of temperature dependent IFCs on the predicted thermal conductivity, we first compare the phonon dispersion and phonon lifetimes computed by different IFCs in Figure 2. As seen in Figure 2(a), with the increase of the temperature, the temperature dependence of frequency spectra is evident. The maximum frequency is reduced from 508 cm$^{-1}$ at 300 K to 480 cm$^{-1}$ at 1500 K. The shift of acoustic branches can also be clearly identified. For example, in Figure 2(b), the frequency change of the lowest TA mode at the X point is shown as a function of the temperature. The dependence agrees with the recent inelastic neutron scattering experiments and numerical simulations.[63] The downshift could be explained by the fact that the shape of the potential surface felt by the atoms at low temperatures (usually the surface near the local minimum of the potential energy) could be different at higher temperatures. Since the shift of phonon frequencies could affect both the phonon group velocities and strengths of phonon scatterings, the thermal conductivity might be different from that from the force constants at the zero-temperature limit.

Integrating the obtained phonon dispersion and lifetimes, we compute and show the phononic thermal conductivity calculated by the temperature-dependent IFCs in Figure 1(a) as circular symbols. It is found that the two sets of IFCs lead to almost the same results, indicating that the role of phonon renormalization on the thermal conductivity of silicon is minor. This could be understood from the fact that most of heat is conducted through the low-frequency acoustic modes. On the one hand, due to the frequency shift, the group velocities of these acoustic modes are reduced by around 4% when the temperature rises from 100 K to 1500 K. One the other hand, the downshift of the frequency could slightly enhance the phonon lifetime according to a simple analytical analysis, as following. For low-frequency modes, the three-phonon annihilation processes are more important than the decay process. The inverse lifetime of the annihilation processes can be computed through

$$1/\tau_{\mathbf{q}s}^{3ph,annih} = \sum_{s's''} \int d^2 S(\mathbf{q}') \frac{1}{|v_{\mathbf{q}'s'} - v_{(\mathbf{q}+\mathbf{q}')s''}|} \omega_{\mathbf{q}s} \frac{\partial n_{\mathbf{q}'s'}}{\partial \omega_{\mathbf{q}'s'}} |V_3|^2 \qquad (19)$$

where $S(\mathbf{q}')$ is the surface area on which wavevector $\mathbf{q}'$ could make both the momentum and energy conservation for the process $\mathbf{q}s + \mathbf{q}'s' \to (\mathbf{q} + \mathbf{q}' + \mathbf{G})s''$ satisfied. We assume the percentage of frequency shift is identical for all modes. While this assumption is not rigorous, it can greatly simplify the analysis. This assumption results in the fact that $S(\mathbf{q}')$ keeps unchanging if the phonon frequencies shift. To deal with the scattering matrix elements, we adopt the Klemens formula[64], which was recently found to work well for cubic crystals and is given by[65]

$$|V_3(\mathbf{q}s, \mathbf{q}'s, \mathbf{q}''s'')| \propto \frac{\sqrt{\omega_{\mathbf{q}s} \omega_{\mathbf{q}'s'} \omega_{\mathbf{q}''s''}}}{v_0} \qquad (20)$$

where $v_0$ is the phonon group velocity in the Debye model. Recognizing that $\frac{\partial n_{\mathbf{q}'s'}}{\partial \omega_{\mathbf{q}'s'}}$ is inversely proportional to the square of frequency and velocities are in portion to the degree of frequency shift, the lifetime should inversely depend on the change of phonon frequency.

Therefore, phonon renormalization results in two completing outcomes, smaller group velocities and longer phonon lifetimes. According to the scaling analysis, the total phononic thermal conductivity should be slightly reduced compared with the results based on 0 K force constants. The numerical calculations are consistent with this analysis as the thermal conductivity using the temperature dependent force constants is 3% larger than that from the 0 K counterparts. Based on these findings, we expect the phonon renormalization is not important to the heat conduction in silicon, but could be crucial for other materials if these materials are much more anharmonic than silicon where the frequency shift is more significant.

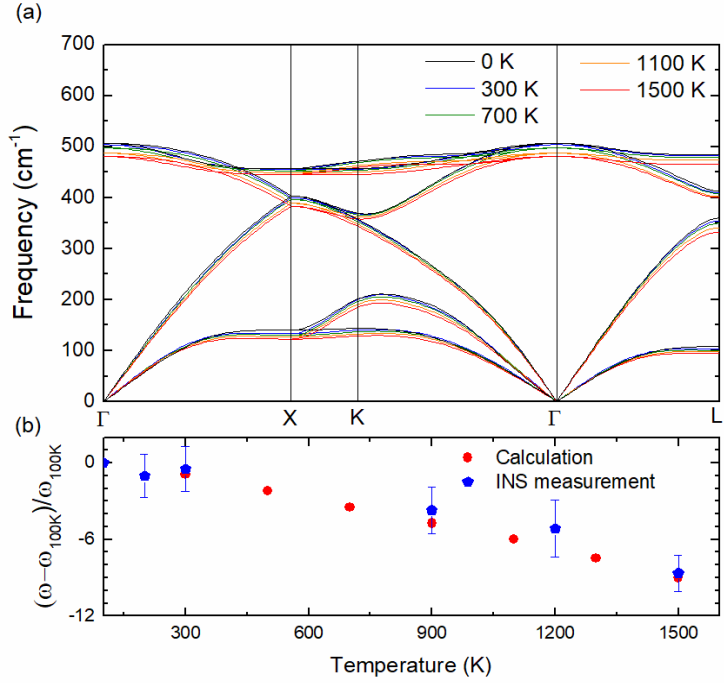

Figure 2. (a) Temperature-dependent phonon dispersion of silicon. (b) The frequency change of the lowest TA mode at the X point as a function of temperature. INS data from measurement is from Ref. [63].

To understand the effects of four-phonon scatterings on the thermal conduction in silicon, the thermal conductivity is computed by considering both three-phonon and four-phonon scatterings, which is plotted in Figure 1(a) as solid triangles. The inclusion of four-phonon scatterings makes the theoretical prediction agree with the measured data up to 1000 K. This range is much more broaden than the three-phonon calculations (below 500 K), clearly indicating that four-phonon scatterings become important when $T > 500\ K$. In fact, even at room temperature, without considering the four-phonon scatterings, the thermal conductivity is overestimated by around 8%. From Figure 1, we can further observe that by including four-phonon scatterings, the thermal conductivity reduces more rapidly with the temperature compared with the three-phonon thermal conductivity in the temperature range we explored here. This phenomenon could be understood from the different temperature dependence of three-phonon and four-phonon lifetimes. Figure 3 shows the phonon lifetimes of Si as a function of phonon frequency at 300 and 1500 K. Temperature alters the relative strength between the three-phonon and four-phonon scatterings. While the four-phonon lifetimes are about ten times larger than the three-phonon ones at 300 K, the four-phonon lifetimes are of magnitude as the three-phonon lifetimes for acoustic phonons at 1500 K. This is because the three-phonon lifetimes are inversely proportional to the temperature while the four-phonon ones exhibit a $T^{-2}$ dependence.[44]

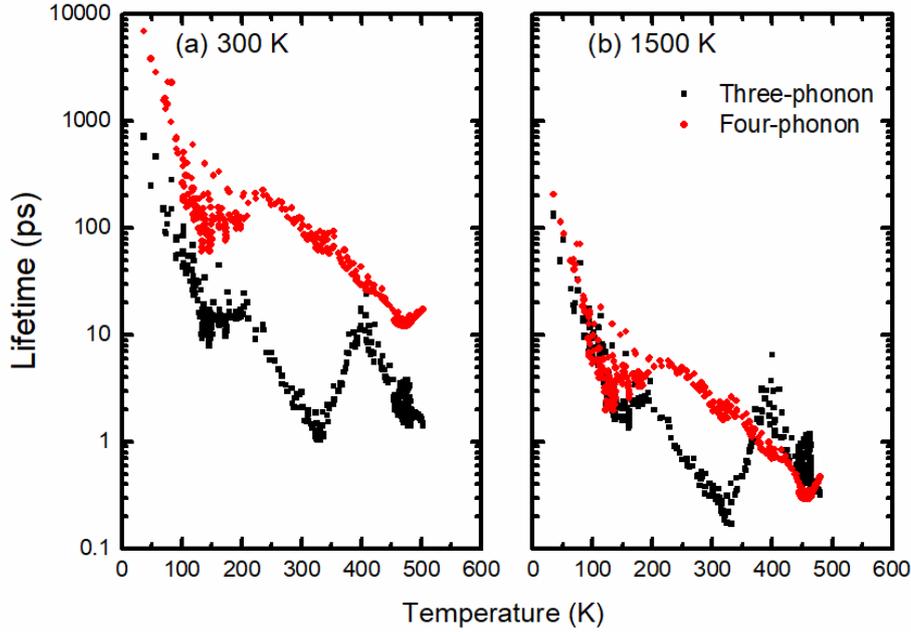

Figure 3. Phonon lifetimes at (a) 300 K and (b) 1500 K as a function of temperature.

We also check the effects of thermal expansion on the phononic thermal conductivity of silicon. We extract the temperature-dependent IFCs at 1500 K by using a lattice constant 0.5% larger than that corresponds to the energy minimization (5.43 Å).[66] The obtained thermal conductivity is only 1.5% lower than the result based on a 5.43 Å lattice constant. The thermal expansion induced change of the thermal conductivity could also be interpreted by the above analysis on the effects of frequency shifting, while keeping in mind that the anharmonic force constants are slightly affected simultaneously. The minor impact of thermal expansion on the thermal conductivity of silicon agrees with previous studies,[22, 67] where the thermal conductivity from a three-phonon RTA calculation.

Although the effects of temperature dependent IFCs, four-phonon scatterings and thermal expansion have been fully taken into account in the modeling, the calculated phononic thermal conductivity of silicon at 1500 K is still considerably smaller than the measured results (*e.g.*, $\kappa^{exp}$ = 22.7 W/mK while $\kappa^{theory}$ = 14.2 W/mK), suggesting the pronounced role of electronic contributions to the thermal conductivity at $T >$ 1000 K, which will be discussed in the following sections.

### B. Electronic thermal conductivity

Figure 4 shows the electronic thermal conductivity calculated from first-principles approach as compared to theoretical models. Both the polar and bipolar components of the electronic thermal conductivity increase significantly with the temperature. This is mainly related to the rapid increase of charge carrier concentration with temperature, which changes from $3.36\times10^{14}$ cm$^{-3}$ at 600K to $9.12\times10^{18}$ cm$^{-3}$ at 1500K. The bipolar component dominates electronic thermal conductivity. The total electronic thermal conductivity reaches 5.09 W/mK at 1500K, which is comparable to the phononic thermal conductivity at the same temperature. While the electronic thermal conductivity is only 0.20 W/mK at 800 K and it can be ignored compared to phononic component. It reveals that the electronic thermal conductivity plays an important role at high temperature range, as it shown in the inset of Figure 6. Glassbrenner and Slack[25] provided the experimental electronic thermal conductivity of silicon in high temperature range. It should be noted the values are not from rigorous direct experimental measurements, but were actually the difference between the total thermal conductivity and extrapolated phononic thermal conductivity. Overall, our calculated total electronic thermal conductivity follows similar trend with experimental results given by Glassbrenner and Slack.[25] The accuracy of experimental result goes in the range of $\pm0.5$W/mK due to the uncertainty in the measurement, as is shown in the purple ribbon of Figure 4.

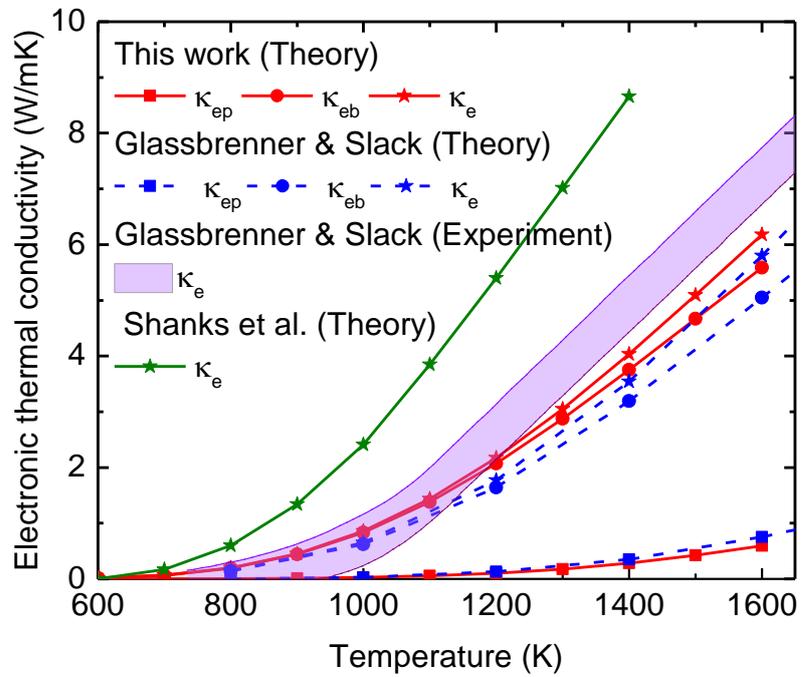

Figure 4. Electronic thermal conductivity of intrinsic silicon: comparison between approximated theoretical results by Shanks *et al.*[62] and Glassbrenner & Slack.[25] The purple ribbon is the experimental

results given by Glassbrenner & Slack,[25] which goes in an accuracy range of ±0.5W/mK. $\kappa_{ep}$, $\kappa_{eb}$, and $\kappa_e$ denotes the polar part, bipolar part, and total electronic thermal conductivity, respectively.

Although there have been several estimations on the electronic thermal conductivity conducted in 1960s, the results are greatly affected by the experimental input parameters. Due to the difficulty of decomposing the electrical transport properties for electron and hole, equation (15) is inconvenient to use to calculate electronic thermal conductivity in previous works. Instead, the following approximated form is widely used[25, 61-62] to estimate the electrical thermal conductivity of semiconductor in high temperature range:

$$k_e = 2\frac{T}{\rho(T)}(\frac{k_B}{e})^2 + \frac{b(T)}{(1+b(T))^2}(4+\frac{\varepsilon_g(T)}{k_B T})^2 \frac{T}{\rho(T)}(\frac{k_B}{e})^2 \quad (22)$$

where $\rho(T)$ is electrical resistance, $b(T)$ is the hole-to-electron mobility ratio and $\epsilon_g(T)$ is the band gap. The first term is polar term and the second term is the bipolar term. Taking experimentally determined resistance, mobility ratio and band gap, Glassbrenner & Slack[25] and Shanks et al.[62] used this equation to estimate the electronic thermal conductivity of intrinsic silicon, and the results are shown in Figure 4 (blue and green curve). The results from the two references do not match with each other because different experimental parameters are used in equation (22). It should be mentioned that this equation is based on the assumption of parabolic electron energy bands. From the comparison with our results and experimental data, it seems that the estimation from Glassbrenner and Slack is more accurate.

Our approach allows to further decompose the electron and hole contribution to the polar electronic thermal conductivity, as shown in Figure 5. Our calculation also shows that the Lorenz number for both electron and holes is around $1.00\times 10^{-8}V^2K^{-2}$, which is close to the value $L = 2k_B^2/(e^2) = 1.48 \times 10^{-8}V^2K^{-2}$ estimated for nondegenerate semiconductors, while far away from the Sommerfeld value $L_0 = \pi^2 k_B^2/(3e^2) = 2.44 \times 10^{-8}V^2K^{-2}$ estimated for metals and degenerate semiconductors. However, since the bipolar contribution is dominant, one cannot directly use Wiedemann-Franz law to estimate the total electronic thermal conductivity for intrinsic semiconductor.

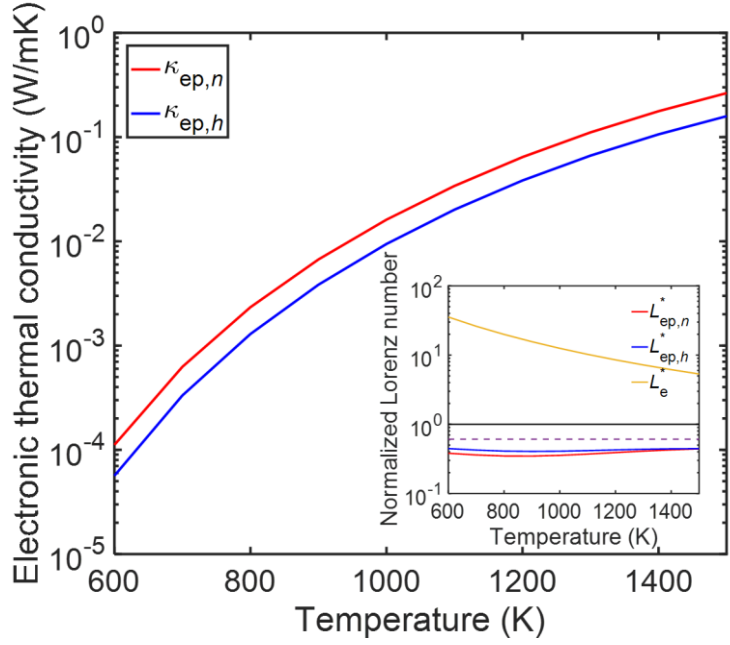

Figure 5. The different electronic thermal conductivity components as a function of temperature. The inset is the Lorenz number. The Lorenz number is normalized by the Sommerfeld value $L_0 = \pi^2 k_B^2/(3e^2) = 2.44 \times 10^{-8} V^2 K^{-2}$. The dash line is theoretical Lorenz number of nondegenerate semiconductors with parabolic electronic bands and acoustic deformation potential approximation, i.e. $L = 2k_B^2/(e^2) = 1.48 \times 10^{-8} V^2 K^{-2}$ and the black line is the Sommerfeld value $L_0$.

### C. Total thermal conductivity

The total thermal conductivities calculated by summing up the contributions from both phonons and electrons, as well as the relative contribution of electrons to the total thermal conductivity, are plotted in Figure 6. The agreement between theory and experiment is reasonably good. At 1500 K, the calculated value is 19.3 W/mK while the measured data by Glassbrenner and Slack is 22.7 W/mK. The slightly lower thermal conductivity values from the calculations should be attributed to the relaxation time approximation, which has been recognized to underestimate the phononic thermal conductivity.[68] While the underestimation could be significant in materials where non-resistive normal phonon-phonon scatterings prevail resistive Umklapp scatterings, previous studies had shown that the relaxation time approximation works reasonably well for silicon at room temperature (the difference from the iterative method is less than 4%).[69] Since under the relaxation time approximation the extrapolation technique[52] to compute the four-phonon lifetimes can be applied and saves the computational costs considerable, we do not attempt to employ the more accurate iterative approach to solve the PBTE, in which the scattering rates for any four-phonon scattering events have to be computed explicitly. In the meantime, the high-temperature thermal conductivity measurements (>1000 K) seem not as reliable as the low-and-middle ones, as the data provided by different groups become scattered at high temperature. Despite the slight discrepancy of the absolute values of the thermal conductivity at high temperature between our theoretical calculations and the measured data, the temperature dependence is faithfully captured by our calculations, confirming that the current framework is capable of predicting the contributions from both phonons and electrons well.

Liao *et al.*[70] found the electron scattering on phonon can significantly decrease the phononic thermal conductivity of heavily doped silicon at room temperature. However, in our calculations on phonon thermal conductivity, we do not include the phonon scatterings caused by electrons. This is attributed to the weak phonon-electron scattering rate. The charge carrier concentration of intrinsic silicon reaches the highest value of ~$9.12\times10^{18}$cm$^{-3}$ at 1500K in our calculation, which is smaller than the concentration where electron can take significant effects. In addition, the 3rd- order and 4th-order phonon-phonon scattering rate at low frequency region is on the high magnitude of ~$10^{-2}$ THz, which is significantly larger than the phono-electron scattering rate (~$10^{-3}$ THz) and greatly weakens the phonon-electron coupling effects.

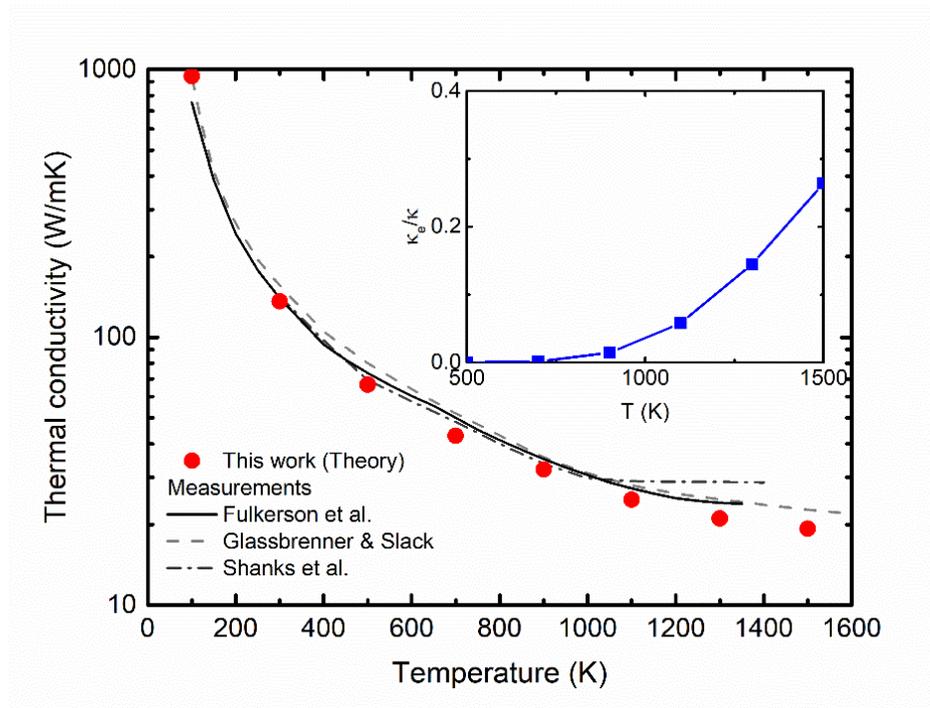

Figure 6. The thermal conductivity as a function of temperature. Our total thermal conductivity value matches well with the results reported by Fulkerson *et al.*[61], Glassbrenner & Slack[25], and Shanks *et al.*[62] The inset is the electronic thermal conductivity contribution to the total thermal conductivity.

## V. Conclusions

In summary, we perform first-principles based Boltzmann transport equation studies to compute both the phononic and electronic thermal conductivity of silicon from 100 K to 1500 K. Compared with conventional three-phonon BTE studies, we combine the effects of four-phonon scatterings, temperature-dependent force constants and thermal expansion when modeling phonon transport. Our calculations demonstrate that four-phonon scatterings result in considerable reduction on the phononic thermal conductivity at $T > 700\ K$. The high temperature induced phonon renormalization leads to flattening of phonon dispersion, but plays minor roles even at 1500 K. Both polar and bipolar contributions to the electronic thermal conductivity is computed through first-principles for the first time. The results confirm that the electronic thermal conductivity due to bipolar effects is one order of magnitude larger than the polar contribution and more than 25% of heat is conducted by charge carriers at 1500 K. Our calculation provides a better understanding on the thermal transport of silicon at elevated temperature range and the calculation scheme can be furtherly expanded to explore the thermal properties of other high-temperature functional materials.

## Acknowledgements

We would like to thank Dr. Zhen Tong for valuable discussions. X. G. acknowledges the support by the National Natural Science Foundation of China No. 51706134. H. B. acknowledges the support by the National Natural Science Foundation of China No. 51676121. Simulations were performed with computing resources granted by HPC (π) from Shanghai Jiao Tong University.